\documentclass{article}
\usepackage{spconf,amsmath,graphicx}
\usepackage[ruled,linesnumbered]{algorithm2e}
\usepackage{multirow}
\usepackage{amssymb,amsmath,bm}
\usepackage{textcomp}
\usepackage{booktabs}
\usepackage[textfont=it,tableposition=top]{caption}
\usepackage{siunitx}
\usepackage{xspace}
\usepackage{url}
\usepackage{lipsum}  
\usepackage{tipa}
\usepackage{enumitem}
\usepackage{pifont}

\newcommand{\xmark}{\ding{55}}

\title{{MBTFNet}: Multi-Band Temporal-Frequency Neural Network For Singing Voice Enhancement}
\name{\begin{tabular}{c}Weiming Xu$^1$, Zhouxuan Chen$^2$, Zhili Tan$^2$, Shubo Lv$^1$, Runduo Han$^1$, \\ Wenjiang Zhou$^2$,  Weifeng Zhao$^2$, Lei Xie$^{1*}$\thanks{* Corresponding author.}\end{tabular}}
\address{$^1$Audio, Speech and Language Processing Group (ASLP@NPU), \\ Northwestern Polytechnical University, Xi'an, China\\
  $^2$Lyra Lab, Tencent Music Entertainment, Shenzhen, China}
\copyrightnotice{979-8-3503-0689-7/23/\$31.00~\copyright2023 IEEE}
\begin{document}
%
\maketitle
\begin{abstract}
A typical neural speech enhancement (SE) approach mainly handles speech and noise mixtures, which is not optimal for singing voice enhancement scenarios where singing is often mixed with vocal-correlated accompanies and singing has substantial differences from speaking. Music source separation (MSS) models treat vocals and various accompaniment components equally, which may reduce performance compared to the model that only considers vocal enhancement. In this paper, we propose a novel multi-band temporal-frequency neural network (MBTFNet) for singing voice enhancement, which particularly removes background music, noise and even backing vocals from singing recordings. MBTFNet combines inter and intra-band modeling for better processing of full-band signals. Dual-path modeling in the temporal and frequency axis and temporal dilation blocks are introduced to expand the receptive field of the model. Particularly for removing backing vocals, we propose an implicit personalized enhancement (IPE) stage based on signal-to-noise ratio (SNR) estimation, which further improves the performance of MBTFNet. Experiments show that our proposed model significantly outperforms several state-of-the-art SE and MSS models.
\end{abstract}
\section{Introduction}
\label{sec:intro}

With easy access to the internet, user-generated content (UGC) has become popular on platforms like TikTok, YouTube, and various karaoke apps. Among UGC, singing recordings provided by users are proliferating. However, these recordings are often accompanied by background noise, reverberation, and accompaniment, as they are recorded in ordinary daily environments. To improve the listening quality and enable further processing such as remixing and singing transcription, the interference needs to be removed.

Recent advances in neural speech enhancement (SE) and music source separation (MSS) can be sensibly leveraged to remove the above interference in the singing recordings. Mask-based time-frequency (TF) domain approaches are prevalent in speech enhancement. In these approaches, a neural network is designed to estimate a \textit{mask} in TF domain from simulated clean-noisy speech pairs. The \textit{mask} is then applied to the noisy signal at runtime to obtain the clean signal. Early approaches only considered the magnitude part of the noisy signal until the complex ratio mask (CRM)~\cite{williamson2015complex} was proposed with explicit consideration of phase. Then complex-valued neural network approaches have become popular for their superior denoising performance. These approaches explicitly model the real and imaginary parts of the speech spectrum typically by a U-net-shaped encoder-decoder structure~\cite{ronneberger2015u, choi2019phase, hu2020dccrn, li2021two, tan2019real}. Research interests have gradually shifted from wide-band (16 kHz) to super-wide-band and full-band~\cite{zhang2022multi, zhao2022frcrn, wang2022harmonicplus}, triggered by the deep noise suppression challenge (DNS) series~\cite{reddy2020interspeech, reddy2021interspeech, dubey2022icassp}. However, increasing the sampling rate leads to a higher modeling complexity. To address this challenge, S-DCCRN~\cite{lv2022s} divides the frequency bands into two parts and performs intra-band and inter-band modeling respectively. MTFAANet~\cite{zhang2022multi} expands the receptive fields of the time-axis and frequency-axis with the specifically designed T-F convolution module (TFCM) to model the challenging full-band signal. HGCN~\cite{wang2022hgcn} and HGCN+~\cite{wang2022harmonicplus} particularly focus on speech harmonics recovery by a harmonic gated compensation network. Recently, personalized speech enhancement (PSE) or target speaker extraction has received a lot of attention~\cite{giri2021personalized,ju2022tea,wang2018voicefilter, ge2020spex+}. In these approaches, enrollment speech from a target speaker can be adopted as a prior feed to the denoising network, leading to superior performance, especially in the case of overlapping speech.

With a similar rationale, the goal of music source separation (MSS) is to particularly separate vocals from background music. In this area, complex-valued U-nets are also dominant~\cite{stoter2019open,defossez2019music,kong2021decoupling,li2021sams,luo2022music,liu2021cws}.
Since the harmonic components that need to be separated between vocals and other instrumental components have a specific frequency band distribution, a sub-band division strategy is usually employed to make the model more focused on a certain frequency band and source type. 
As a typical approach, ResUNetDecouple~\cite{kong2021decoupling} uses a very deep structure, a residual UNet architecture with up to 143 layers, and achieves state-of-the-art separation performance on the popular MUSDB18~\cite{rafii2017musdb18} dataset.

Although both speaking and singing originate from the same human vocal system, they have substantial differences in phoneme usage, tonality, diction, breathing, and volume. For example, singing has a higher average intensity level than speech and always features a wider intensity variation than speech. Likewise, singing occurs at higher frequency levels than speaking and within a wider range of frequencies. Differently in pronunciation with speaking, in singing, we always extend the vowels to the greatest length possible because they carry most of the sound, but consonants are usually shortened as they are much harder to project. Singing has a specific rhythm and melody to adhere to. Sustained notes and vibrato differentiates it from speech. Sometimes in singing, the lead vocal is also accompanied by backing vocals. On the other hand, the background music associated with singing also has unique characteristics. As a coherent background, musical accompaniments mostly are harmonic, broadband, and highly correlated with singing.

In this paper, we present a neural network approach designed specifically for enhancing singing voices. Our goal is to remove musical accompaniments, various types of noise, and even backing vocals. Our work is inspired by the recent advances in SE and MSS reviewed above, but it makes substantial improvements that target the unique characteristics of singing. Specifically, we propose a novel multi-band temporal-frequency neural network (MBTFNet) with the following designs:

\begin{itemize}
    \item We design an inter-band and intra-band modeling structure to make it easier in distinguishing harmonic structures of vocals and background music at the \textit{frequency} scale. 
    \item To better distinguish fine-grained harmonic structures between vocals and background music at a \textit{temporal} scale, we introduce time-axis dilation block (TDB),  dual-path RNN (DPRNN)~\cite{luo2020dual} and squeezed-TCM (STCM)~\cite{li2021two} to expand the receptive field of the singing enhancement model.
    \item Inspired by the recent advances in personalized speech enhancement, we propose an implicit personalized singing enhancement module as a secondary stage to further remove residuals and backing vocals. Using a signal-to-noise ratio (SNR) estimator, the module can dynamically leverage the singer's singing as a speaker embedding without requiring explicit voice enrollment from the singer.

\end{itemize}

\noindent Experiments show that the proposed MBTFNet outperforms several state-of-the-art SE and MSS models in singing voice enhancement by a large margin. With the help of the implicit personalized enhancement module, a further performance gain can be obtained including the challenging case for backing vocal removal.

\section{MBTFNet}
\label{sec:format}

\begin{figure*}[htbp]
    \centering 
    \includegraphics[width = 16cm]{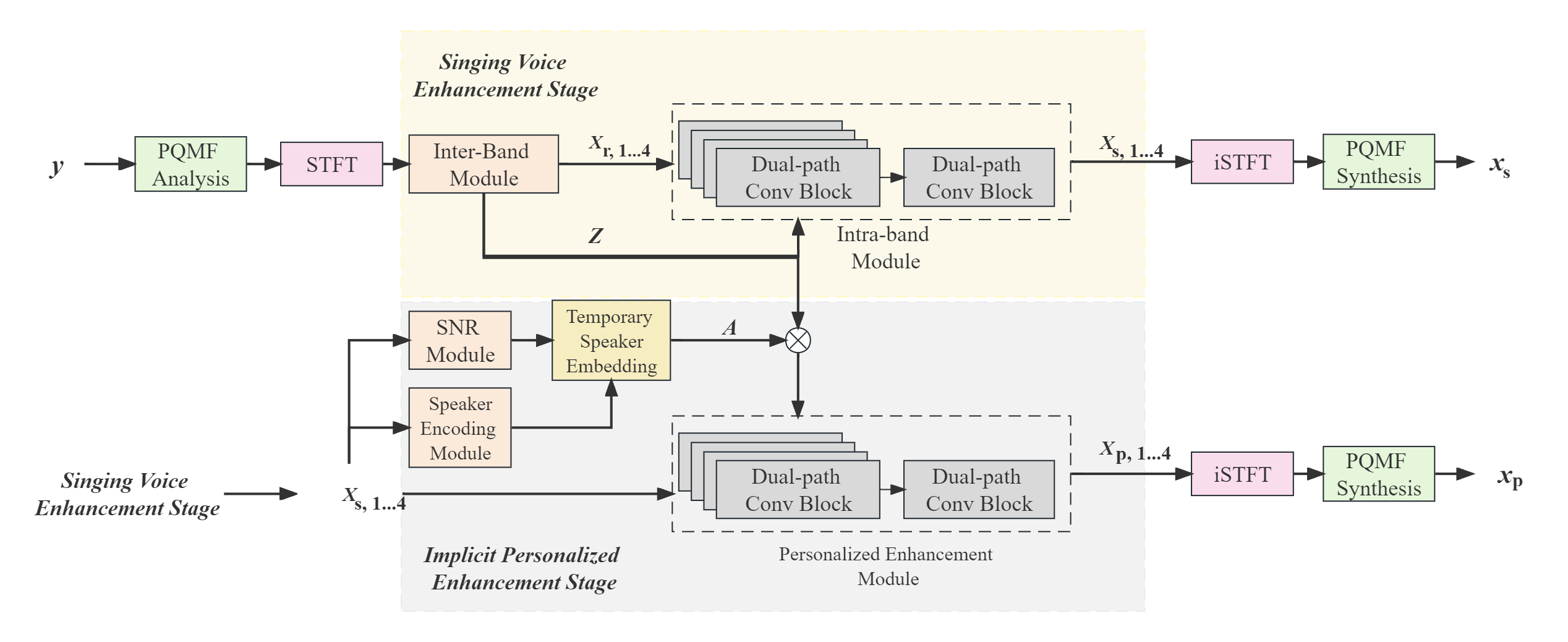}
    \caption{\text{The overall network structure of MBTFNet.}}
    \label{fig:mbtfnet_main}
\end{figure*}

The noisy singing signal can be described as:
\begin{equation}
    Y(t,f)=X(t,f)+N(t,f)+M(t,f)+B(t,f)
\end{equation}
where $Y(t,f)$, $X(t,f)$, $B(t,f)$, $M(t,f)$ and $N(t,f)$ represent the noisy signal, clean singing voice, backing vocal voice, background music and noise TF-bins, respectively. In our scenario, the goal is to extract the clean singing voice $X(t,f)$ from the input noisy signal $Y(t,f)$. In the singing voice enhancement (SVE) stage, we introduce the multi-band temporal-frequency neural network (MBTFNet), consisting of both inter and intra-band modeling, to remove the background music $M(t,f)$ and noise $N(t,f)$ from the input signal $Y(t,f)$. For the rest part, we further eliminate the backing vocal signal $B(t,f)$ by an implicit personalized enhancement (IPE) stage. Fig.~\ref{fig:mbtfnet_main} shows the overall structure of MBTFNet with SVE and IPE stages.

\begin{figure*}[htbp]
    \centering 
    \includegraphics[width = 16cm]{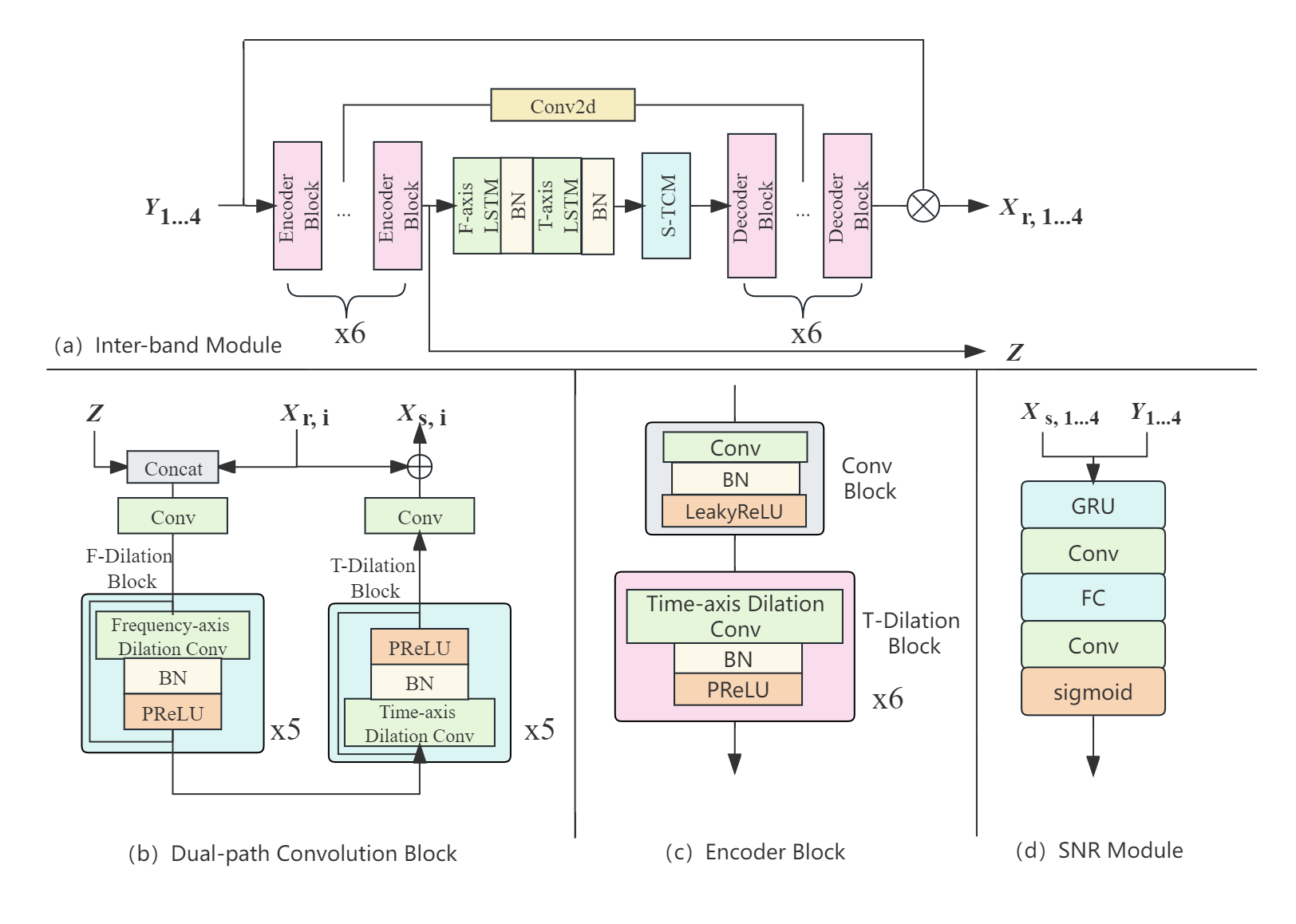}
    \caption{The design of the inter-band module (a), the dual-path convolution block (b), the encoder block (c), and the SNR module (d).} 
    \label{fig:mbtfnet_detail}

\end{figure*}

\subsection{Inter- and Intra-Band Modeling}

Different from noise, background music has lots of harmonic structure correlated with vocals, which causes difficulty in singing voice enhancement. 

The complexity of directly modeling full-band signals is large, and the harmonic components of musical instrument are often distributed within a specific frequency band. So we design an inter-band and intra-band modeling structure. Meanwhile, larger receptive field modules are introduced to better distinguish harmonic structures between vocals and music. The overall process is shown in Fig.~\ref{fig:mbtfnet_main}, where the input audio signal $y$ is first decomposed into $C$ sub-band signals using Pseudo Quadrature Mirror Filter (PQMF)~\cite{Vaidy1993pqmf}. The complex-valued spectrogram $Y_i \in \mathbb{C}^{F\times T}$, $i=1, ..., C$ are computed by STFT, where $F$ and $T$ are the frequency and time index, and then fed into the inter-band module to get the rough enhanced result, $X_{r,i}$. Similarly to general speech enhancement models, this module learns the common characteristic of different sub-bands. Specifically, it adopts the U-Net structure, as shown in Fig.~\ref{fig:mbtfnet_detail}(a). The encoder block consists of a conv block and multiple stacked time-axis dilation blocks (TDB) which help to expand the receptive field. TDB consists of a convolution layer with time-axis dilation, a batch norm layer, and a PReLU layer, as shown in Fig.~\ref{fig:mbtfnet_detail}(c). The output of the stacked encoder blocks, $Z\in \mathbb{C}^{N\times K\times T}$, contains the full-band information and thus will be used in the subsequent intra-band module and personalized enhancement module, where $N$ and $K$ are transformed from $C$ and $F$ by encoder blocks. The decoder block has a similar structure to the encoder, where the convolution layer in the conv block is replaced by a transpose convolution layer. For the stacked TDBs, the higher one has a higher number of dilation, in order to make the higher layer have a larger receptive field. Specifically speaking, the number of dilation is $2^n$, where $n$ is the layer index beginning from 1.

In order to fully utilize the full-band information, each rough enhanced sub-band $X_{r,i}$ is concatenated with the full-band feature $Z$, as the input of dual-path convolution blocks (DPCB). The intra-band module aims to learn the unique characteristic of each band, thus it consists of $C$ DPCBs. Each DPCB focus on the harmonic components of musical instrument in its corresponding sub-band, which makes the learning process easier and the elimination ability better. DPCB is stacked by two types of layers, namely the time-axis dilation convolution layer and the frequency-axis dilation convolution layer, which helps to extract features of each sub-band adequately, as shown in Fig.~\ref{fig:mbtfnet_detail}(b). The higher dilation block has a higher number of dilation, as the $2^n$ setup above.

After removing the background music and noise in each band, another DPCB is used to merge the enhanced sub-bands as $X_{s,i}$, as shown in Fig.~\ref{fig:mbtfnet_main}.

\subsection{Implicit Personalized Enhancement}

To further eliminate the backing vocals and the residual harmonics of background music, we design an implicit personalized enhancement (IPE) stage to extract the voice of the lead vocalist. It consists of an SNR module~\cite{lv2021dccrn+}, a speaker encoding module (SEM) and a personalized enhancement module (PEM). The SNR module is responsible for evaluating the cleanliness of $Y(t,f)$, and the SEM extracts the speaker embedding from the previous enhancement result $X_{s,i}$. The PEM further removes the backing vocals and the residual harmonics of background music according to the cleanliness and the speaker information above.

The SNR module is shown in Fig.~\ref{fig:mbtfnet_detail}(d), which is composed of a GRU layer, a convolution layer, and a sigmoid layer. The SEM is a pre-trained ECAPA-TDNN~\cite{desplanques2020ecapa}. The PEM has roughly the same structure as the intra-band module, and an additional convolution layer is added to map the output $a\in \mathbb{R}^{192}$ of SEM to $A\in \mathbb{R}^{N\times K}$. $A$ is multiplied by $Z$ and combined with $X_\text{s}$ as the input of PEM to get $X_\text{p}$.


Temporary speaker embedding $E$ is updated using the SNR module and SEM. The SNR module aims to estimate the cleanliness score $S$ of $y$. The cleanliness score $S$ at time $t$, $S_t$, can be obtained from $X$ and $Y$ as $ S_t= \log_{10} ({ \sum_{f} |X_{t,f}| }/{ \sum_{f}|Y_{t,f}|})$ and it is normalized by its mean and variance in the entire training set.
 
For inference, we update the temporary speaker embedding when the model predicted cleanliness score $\hat{S}$ is high enough. A higher $\hat{S}$ means a more reliable speaker embedding. The update rule is shown in Algorithm~\ref{algorithm:update_e}. The weight $\alpha$ and SNR threshold $\lambda$ in Algorithm~\ref{algorithm:update_e} are hyper-parameters. 

\begin{algorithm}
    \KwIn{$X_s, \lambda$}
    \KwResult{$E$}
    $E \gets \vec{1}$\;
    \While{chunk in chunks}{
    $\Bar{\hat{S}} \gets \frac{1}{T}\sum_{t} \text{SNR}(X_{s,t}, Y_{t})$\;
    \If { $\Bar{\hat{S}} >= \lambda$}{
        $ E \gets \alpha E + (1-\alpha)\text{SEM}(X_s) $\;
        }
    }
    \caption{Update temporary speaker embedding $E$ in inference.}
    \label{algorithm:update_e}
\end{algorithm}

During training, we use a clean enrollment to estimate the speaker embedding, therefore we assume it is always clean enough and do not require SNR threshold $\lambda$. The hyper-parameter $\lambda$ for inference can be set manually. In our experiments, it is learned as the mean of $\Bar{\hat{S}}$ in Algorithm~\ref{algorithm:update_e} during training and referring to it as $\lambda_t$.
 

\section{Training Objective}

For the SVE stage, we combine the SI-SNR~\cite{le2019sdr} loss $\mathbb{L}_{\text{SI-SNR}}$ with complex mean square error (cMSE) loss $\mathbb{L}_{\text{cMSE}}$ as the loss function: $ \mathbb{L}_{\text{SVE}} = \mathbb{L}_{\text{SI-SNR}}(X, X_s) + \mathbb{L}_{\text{cMSE}}(X, X_s)$.

For the IPE stage training, the frozen well-trained inter and intra-band modules of MBTFNet are required. Then the model needs to learn both SNR estimation and personalized enhancement. The SNR score loss $\mathbb{L}_{\text{SNR}}$ is defined as the mean square error (MSE) between the ground truth $S$ and the model predicted score $\hat{S}$.

The final loss function of the IPE stage is:

\begin{equation}
\begin{split}
    \mathbb{L}_\text{IPE} &= \mathbb{L}_{\text{SI-SNR}}(X, X_p) + \mathbb{L}_{\text{cMSE}}(X, X_p) \\
    &+ 10\mathbb{L}_{\text{SNR}}(S, \hat{S}).
\end{split}
\end{equation}

\section{Experiment}
\subsection{Datasets}

We utilize the clean vocals and accompaniments from the MUSDB18HQ~\cite{rafii2017musdb18} dataset, which consists of 100 training tracks and 50 test tracks. For our experimental noise set, we employ the noise set of Deep Noise Suppression 2022 (DNS2022)~\cite{dubey2022icassp} and randomly select 60 hours for training noise and 10 hours for test noise.


The training pairs are simulated by dynamically mixing the vocal and its corresponding accompaniment with a random SNR ranging from -5 to 15dB, and then mixing the mixture and noise with a random SNR ranging from -5 to 15 dB. To simulate the test set, we apply the same rules to each vocal track of the MUSDB18HQ test set. Each test vocal track is used to simulate 5 mixture audio, resulting in a total of 250 audio tracks.

MUSDB18HQ is a classic music separation dataset, but some of its vocal tracks contain both lead vocals and backing vocals. This can cause issues during the training and testing of the IPE stage. Therefore, we use the M4Singer~\cite{zhangm4singer} dataset for the IPE stage experiments. M4Singer is an unaccompanied singing dataset that includes 10 male and 10 female singers, with a total of 700 vocals. Vocals from 8 males and 7 females are used in the training, while the remaining vocals are used for testing. We apply the same rules as the MUSDB18HQ test set to generate the without-backing test set. We then mix the without-backing test set with vocals to generate the random-backing test sets. The random-backing test sets select random vocals as the backing vocals, which can result in gaps with the real audio. To address this, we refer to~\cite{yuan2022improved} and propose a selection method to simulate the selected-backing test set. We randomly select 10 vocal tracks and choose the one with the highest cross-correlation of chroma features with the lead vocal, then randomly raise or lower the backing vocal by two semitones, and then mix them with a random SNR ranging from 5 to 10 dB to generate desired audio. 

\subsection{Training Setup and Baselines}
All training and test data are at a 44.1kHz sampling rate. A frame of 20 ms with a shift of 10 ms is used for STFT computation. We vary the learning rate during training as
\begin{equation}
    lr=d^{-0.5} \cdot \min(\text{step}^{-0.5}, \text{step} \cdot \text{warmup\_steps}^{-1.5}) 
\end{equation}
where $d=1\mathrm{e}{-3}$ and warmup\_steps=5000. All models in Table \ref{tab:musdb} were trained using the Adam optimizer under identical conditions to suppress both noise and music until no further improvement was observed. The detailed configuration of MBTFNet is as follows, and the other models for comparative experiments on MUSDB18HQ are configured according to their papers. The SEM uses a pre-trained ECAPA-TDNN model and fixes the weights in experiments.

The PQMF splits the signal into 4 sub-band signals. The inter-band module has 6 encoder blocks and decoder blocks, the channel of encoder blocks are [8, 64, 64, 64, 128, 128], and the kernel size of encoder blocks are (5, 2). Each encoder block contains 6 TDBs and all the kernel size in TDB is (3,3). The decoder block is the same as the encoder block.  The layer number of the frequency-axis and time-axis is 2, and their rnn-units are 256 with 1 STCM layer. The intra-band module consists of 4 DPCBs. Each DPCB has 5 FDBs and 5 TDBs, and their kernel size is (3,3).

The ECAPA-TDNN configures with 1024 channels. In the SNR Module, the number of GRU layers is 2 and the rnn-units are 256, followed by a convolution layer with an input channel of 1, an output channel of 2, and a kernel size of (3, 3). The DPCB in PEM is the same as the intra-band module.

When conducting experiments on the M4Singer dataset, we
first train the SVE part for 100 epochs, then freeze it to train the IPE part for 50 epochs. When training the IPE part, the mixture has a backing vocal with a probability of 0.5, and corresponding enrollment audio is provided.

\subsection{Experimental Results and Discussion}

We first conduct model comparison and ablation experiments on the MUSDB18HQ simulation test set, using SI-SNR and PESQ as objective metrics, and the experimental results are shown in Table~\ref{tab:musdb}. MBTFNet (causal) achieves the highest metrics among experiment speech enhancement models, where SI-SNR and PESQ are 1.39dB and 0.14 higher than the second-best S-DCCRN, respectively. MBTFNet (no causal) is better than the music separation model ResUNetDecouple, where SI-SNR and PESQ are 1.63dB and 0.32 higher, respectively. 

MBTFNet-A, MBTFNet-B and MBTFNet-C are ablation experiments, where MBTFNet-A means that the full-band signal is directly modeled without PQMF, MBTFNet-B means that all sub-bands are modeled in one Intra-band Module, 
and MBTFNet-C means that the Intra-band Module is removed. We keep the parameters of MBTFNet-A and MBTFNet-B consistent with MBTFNet. Ablation experiments show that modeling each sub-band individually improves the performance of MBTFNet.
\begin{table}[htbp]		
\caption{Comparison with various models on MUSDB18HQ simulation test set.}
\centering
\label{tab:musdb}
\resizebox{\linewidth}{!}{
    \begin{tabular}{lcccc}
    \toprule
    Model           & Causal      & Param. (M) & SI-SNR (dB) & PESQ \\
    \midrule
    Noisy           & -           & -          & 0.49        & 1.79 \\
    HGCN+~\cite{wang2022harmonicplus}           & $\checkmark$     & 7.03       & 7.25        & 2.49 \\
    S-DCCRN~\cite{lv2022s}         & $\checkmark$     & 2.04       & 7.43        & 2.60 \\
    MBTFNet         & $\checkmark$     & 4.08       & \textbf{8.82} & \textbf{2.74} \\
    ResUNetDecouple~\cite{kong2021decoupling} & \xmark           & 103        & 7.97        & 2.63 \\
    MBTFNet-A      & \xmark           & 8.57       & 8.62        & 2.80 \\
    MBTFNet-B      & \xmark           & 8.54       & 8.49        & 2.86 \\
    MBTFNet-C      & \xmark           & 8.43       & 9.11        & 2.94 \\
    MBTFNet         & \xmark           & 8.54       & \textbf{9.60} & \textbf{2.95} \\
    \bottomrule
    \end{tabular}
}
\end{table}

After verifying the model performance of MBTFNet on the MUSDB18HQ simulation test set, we continue to experiment with the IPE on the M4Singer simulation test set. The experiment tests three values of $\lambda$: 0, $\lambda_t$, and 1. When $\lambda$=0, every $X_s$ is accepted to update the speaker embedding. When $\lambda$=1, $X_s$ is not accepted to update the speaker embedding, making two-stage MBTFNet degenerate into one-stage MBTFNet. $\lambda_t$ is obtained during training. The experiments are conducted on three types of test sets: without-backing, selected-backing, and random-backing. The results are shown in Table~\ref{tab:m4singer}. Where Noisy, IPE, and PE respectively represent no enhancement, implicit personalized enhancement, and personalized enhancement.
First, we compare the IPE at different values of $\lambda$. Among all $\lambda$ values tested, $\lambda_t$ achieves the highest metrics in all three test sets. Compared to $\lambda$=1, $\lambda$=$\lambda_t$ achieves a further improvement on the test set. This demonstrates that using the IPE can further remove residual noise generated by the SVE stage. $\lambda$=0 receives the lowest metrics, which indicates that the SNR estimator contributes to the IPE stage.

Second, we compare the performance of temporary speaker embedding with directly provided speaker enrollments obtained from the same singer's other songs, which we refer to as personalized enhancement (PE). The explicit speaker enrollment is extracted from a random 20-second segment by the same singer in the test set. As shown in Table~\ref{tab:m4singer}, IPE has better performance than PE. The reason is that the speaker module designed for speech undergoes a reduction in its effectiveness when applied in the singing scene. In the case of PE, there may be a mismatch between the extracted speaker characteristics from the enrollment and test audio, as they are obtained from different songs, leading to a decrease in the performance of personalized enhancement. In our proposed IPE module, the temporary speaker embedding is extracted from the previous part of the test audio, thus the speaker characteristic suffers from less mismatch.

\begin{figure}[htbp]
\centering

\begin{minipage}{0.32\linewidth}
    \centering
    \includegraphics[width=0.9\linewidth]{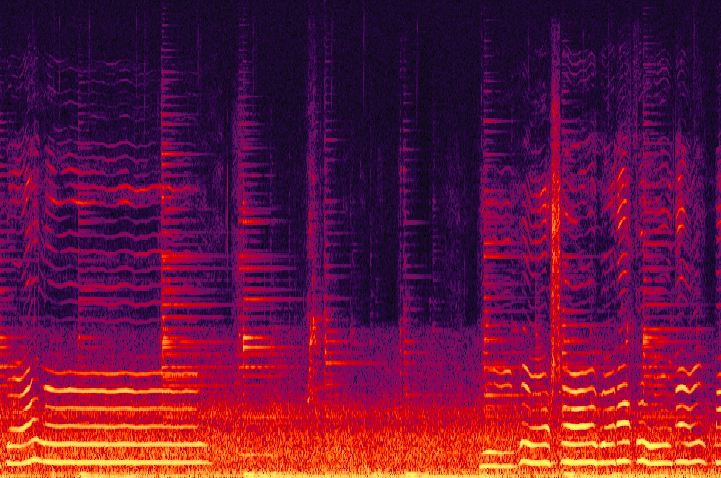}
\end{minipage}
\begin{minipage}{0.32\linewidth}
    \centering
    \includegraphics[width=0.9\linewidth]{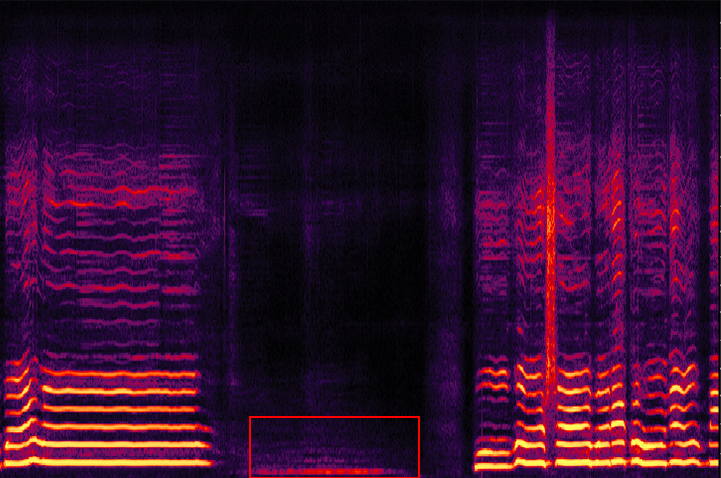}
\end{minipage}
\begin{minipage}{0.32\linewidth}
    \centering
    \includegraphics[width=0.9\linewidth]{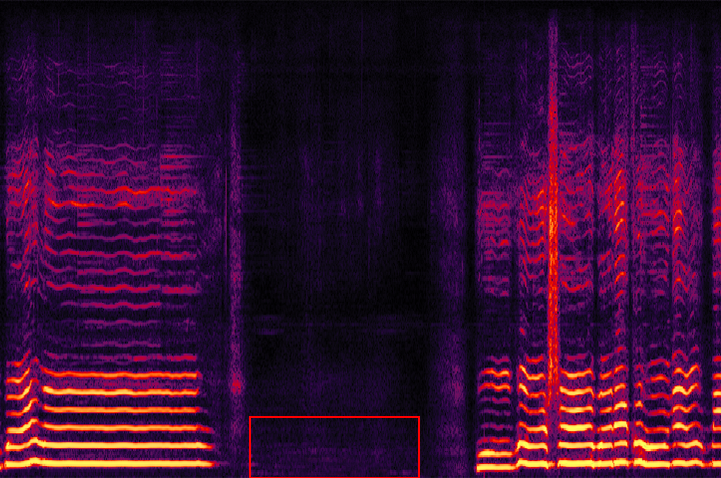}
\end{minipage}

\begin{minipage}{0.32\linewidth}
    \centering
    \includegraphics[width=0.9\linewidth]{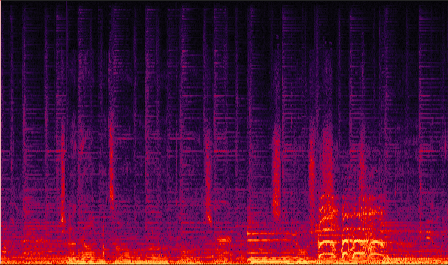}
\end{minipage}
\begin{minipage}{0.32\linewidth}
    \centering
    \includegraphics[width=0.9\linewidth]{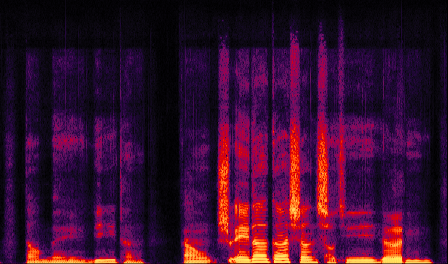}
\end{minipage}
\begin{minipage}{0.32\linewidth}
    \centering
    \includegraphics[width=0.9\linewidth]{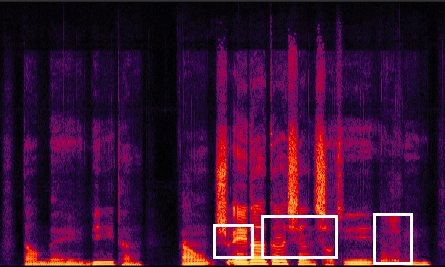}
\end{minipage}

\caption{MBTFNet samples on the without-backing (above) and selected-backing (below) test set. The input audio (left) contains noise, accompaniment, etc. The output of the SVE stage (middle) only has some accompaniment residuals, and the output of the IPE stage (right) further removes them.}
\label{img:compare}
\end{figure}

 
Fig.~\ref{img:compare} shows the spectra of $\lambda$ valued 1 and $\lambda_t$ in the without-backing (above) and selected-backing (below) test set. In the without-backing samples, there are some piano accompaniment residuals if only using the SVE stage, but these residuals are further eliminated in the IPE stage. In the second group of samples (below), the backing vocalist is eliminated by the IPE stage.
\begin{table}[htbp]
\caption{Performance on personalized singing voice enhancement for MBTFNet on M4Singer simulation test set. When $\lambda=1$, no speaker embedding is updated by SEM; when $\lambda=\lambda_t$, the speaker embedding is automatically updated if the enhanced speech is clean enough; when $\lambda=0$, the speaker embedding is always updated.}
\centering
\label{tab:m4singer}
\resizebox{\linewidth}{!}{
\begin{tabular}{lcccccccc}
\hline
 & \multirow{2}{*}{Use SEM} & \multirow{2}{*}{$\lambda$} & \multicolumn{2}{c}{without-backing} & \multicolumn{2}{c}{selected-backing} & \multicolumn{2}{c}{random-backing} \\
 &  &  & SI-SNR & PESQ & SI-SNR & PESQ & SI-SNR & PESQ \\ \hline
Noisy &  & - & 0.56 & 1.17 & -1.28 & 1.13 & -1.23 & 1.15 \\
IPE & \xmark & 1 & 14.17 & 3.06 & 7.54 & 2.60 & 7.48 & 2.59 \\
IPE & Auto & $\lambda_t$ & \textbf{14.50} & \textbf{3.07} & \textbf{7.95} & \textbf{2.68} & \textbf{8.01} & \textbf{2.69} \\
IPE & $\checkmark$ & 0 & 13.13 & 3.00 & 6.92 & 2.53 & 6.98 & 2.52 \\
PE & $\checkmark$ & - & 13.14 & 2.93 & 7.73 & 2.62 & 7.82 & 2.67 \\ \hline
\end{tabular}
}
\end{table}

\section{Conclusions}
This paper has focused on singing voice enhancement, designing a novel Multi-Band Temporal-Frequency Neural Network (MBTFNet) to particularly address the challenges in the singing scene. By further introducing an implicit personalized enhancement (IPE) stage with an automatic speaker enrollment strategy, MBTFNet gets a stronger ability to distinguish the target singer from background music and backing vocals. Our proposed model achieves the best metrics in the MUSDB18HQ simulation test set as compared with SOTA SE and MSS models and obtains significant improvement on the M4singer simulation test set by using the IPE stage. 

\bibliographystyle{IEEEbib}
\bibliography{main}

\end{document}